\documentclass[preprint, prb, showpacs, superscriptaddress]{revtex4}
\usepackage{epsfig}
\begin{document}
\title {
Electronic Structure and Optical Properties of the Co-doped
Anatase TiO$_{2}$ Studied from First Principles }

\author {Hongming Weng}
\affiliation {Group of Computational Condensed Matter Physics,
National Laboratory of Solid State Microstructure and Department
of Physics, Nanjing University, Nanjing 210093, People's Republic
of China}

\author {Xiaoping Yang}
\affiliation {Group of Computational Condensed Matter Physics,
National Laboratory of Solid State Microstructure and Department
of Physics, Nanjing University, Nanjing 210093, People's Republic
of China}
\affiliation{Department of Physics, Huainan Normal
University, Huainan, Anhui 232001, People's Republic of China}

\author {Jinming Dong}
\affiliation {Group of Computational Condensed Matter Physics,
National Laboratory of Solid State Microstructure and Department
of Physics, Nanjing University, Nanjing 210093, People's Republic
of China}
\author {H. Mizuseki}
\author {M. Kawasaki}
\author {Y. Kawazoe}
\affiliation {Institute for Materials Research, Tohoku University,
Sendai 980-8577, Japan}

\date{\today}

\begin{abstract}
The Co-doped anatase TiO$_{2}$, a recently discovered
room-temperature ferromagnetic insulator, has been studied by the
first-principles calculations in the pseudo-potential plane-wave
formalism within the local-spin-density approximation (LSDA),
supplemented by the full-potential linear augmented plane wave
(FP-LAPW) method. Emphasis is placed on the dependence of its
electronic structures and linear optical properties on the
Co-doping concentration and oxygen vacancy in the system in order
to pursue the origin of its ferromagnetism. In the case of
substitutional doping of Co for Ti, our calculated results are
well consistent with the experimental data, showing that Co is in
its low spin state. Also, it is shown that the oxygen vacancy
enhances the ferromagnetism and has larger effect on both the
electronic structure and optical properties than the Co-doping
concentration only.
\end{abstract}

 \maketitle

\section{introduction} \label{introduction}
The discovery of the ferromagnetic semiconductor has stimulated a
great deal of interests in the origin of its ferromagnetism
because of its potential applications in the spintronics, a
rapidly developing research area, in which the electron spin plays
an important role in addition to the usual charge degree of
freedom. In fact, the diluted ferromagnetic semiconductors (DMSs)
based on II-VI compound semiconductor had been studied for over
twenty years~\cite{1}. The transition-metal magnetic impurities
are often used as spin injector to cause the giant
magneto-resistance and magneto-optical effect. But usually, it is
the n-type carrier doping with concentration of 10$^{19} cm^{-3}$
in maximum and the p-type doping is difficult. This drawback
results in a lower ferromagnetic Curie temperature T$_{c}$ in the
order of $\sim$1 K, which is well explained by the carrier-induced
mechanism~\cite{2}. On the other hand, ferromagnetism with Curie
temperature as high as 110 K was discovered~\cite{3} in III-V
compound based DMSs, such as GaAs doped with Mn ion. The strong
$p-d$ exchange interaction intermediated by mobile holes is
thought as the origin of ferromagnetism with so higher Curie
temperature~\cite{4}. However, both of them are still far from the
practical use at room-temperature. Therefore, a lot of efforts has
been made continuously in the recent years in order to get higher
and higher T$_{c}$ DMSs. Motivated by this, the transition metal
oxide based DMSs have been buoyed up as a candidate of the
room-temperature DMSs.

Excitedly, Matsumoto {\it et al.}~\cite{5} reported that Co-doped
anatase TiO$_{2}$ (Ti$_{1-x}$Co$_{x}$O$_{2 -\delta})$ can keep
ferromagnetic order up to 400 K with magnetic moment about 0.32
$\mu_{B}$/Co, which was explained by the carrier-induced
ferromagnetism with the exchange interaction mediated by electrons
not holes. Chamber {\it et al.}~\cite{6,7} had reproduced the
ferromagnetic Ti$_{1 - x}$Co$_{x}$O$_{2 - \delta }$, but found the
magnetic moment on Co is as high as 1.25 $\mu _{B}$. They
considered that the ferromagnetism strongly depends on the oxygen
deficiency. Because recent theoretical~\cite{8} and
experimental~\cite{9} studies show that oxygen vacancies resulted
from the substitution of Co (II) for Ti(IV) do not contribute to
carriers, J. Y. Kim {\it et al.}~\cite{10} and S. R. Shinde {\it
et al.}~\cite{11} paid more attention to the role of Co in the
origin of the ferromagnetism. They thought that some amount of Co
clusters induce the ferromagnetism, but their thermal treatment is
quite questionable. And in fact, the measurement of magnetic
circular dichroism (MCD) spectrum done by T. Fukumura ``rules out
the apparent precipitation of Co metal as a source of
ferromagnetic signal"~\cite{12}. More recently, S. B. Ogale {\it
et al.}~\cite{13} found that the thin films of
Sn$_{1-x}$Co$_{x}$O$_{2-\delta}$ (x$<$0.3) not only exhibit
ferromagnetism with a Curie temperature close to 650 K, but also a
giant magnetic moment of 7.5$\pm$0.5 $\mu_{B}$/Co, which is much
larger than the value for small Co clusters ($\sim$2.1
$\mu_{B}$/Co) and offers another evidence to disfavour the Co
cluster contribution to the ferromagnetism in the transparent
Co-doped TiO$_{2}$ and SnO$_{2}$. J. R. Simpson {\it et
al.}~\cite{14} studied the optical conductivity of the
Ti$_{1-x}$Co$_{x}$O$_{2-\delta}$, and concluded that the Co atom
is not substitutional but interstitial or forms Co-Ti-O complex
based upon the absence of in-band-gap absorption and the blue
shift of absorption edge with increasing ${x}$. On the other hand,
the first-principles calculation done by James M.
Sullivan~\cite{8} supported this idea, and showed that ``n-type
behavior in Co-doped TiO$_{2}$ requires a substantial amount of Co
to be in the interstitial form, and that this can only happen
under O-poor growth conditions''. So, at the present time, the
origin of the ferromagnetism and the high Curie temperature of
Ti$_{1 - x}$Co$_{x}$O$_{2-\delta}$ are still in controversial, and
more experimental and theoretical studies are needed.

In this paper, we have used CASTEP~\cite{15} to study the
electronic structures and optical properties of
Ti$_{1-x}$Co$_{x}$O$_{2 -\delta}$ in four different
configurations, which are (x=0.0417, $\delta$=0), (x=0.0625,
$\delta$=0) and (x=0.0625, $\delta$=0.0625) with oxygen vacancy
near Co atom and Ti atom, respectively. Our results show that
variation of x from 0.0417 to 0.0625 causes little difference of
the electronic structures and optical properties, but the oxygen
vacancy changes them greatly. In the local spin density
approximation (LSDA), the magnetic moment about 0.68 $\mu_{B}$ per
Co atom has been obtained in the first and second case. The oxygen
vacancy near Co atom in the third case causes increasing of the
magnetic moment on Co to about -0.90 $\mu_{B}$, while in the
fourth case the oxygen vacancy near Ti changes this value to -0.38
$\mu_{B}$. The relative positions of the optical conductivity
peaks in the first three configurations are quite comparable with
the experimental results~\cite{14}, while in the fourth one they
are totally different. All these results support the idea of
Chamber {\it et al.}~\cite{6,7} that the ferromagnetism strongly
depends on the oxygen deficiency. Moreover, total energy of the
system in the third case is lower than that in the fourth one by
about 0.33 eV per anatase cell, indicating that the oxygen vacancy
prefers to stay near Co than near Ti~\cite{9}.

\section{Calculation Method} \label{Calculation Method}
The software package CASTEP~\cite{15} has been used in our
calculations, which is based on a total energy pseudo-potential
plane-wave method within the local spin density approximation
(LSDA). In the calculation, the Perdew and Zunger's form of the
exchange-correlation energy has been used~\cite{16,17}. The
ion-electron interaction is modelled by ultra-soft local
pseudo-potentials in the Vanderbilt form~\cite{18}. The maximum
plane wave cut-off energy is taken as 280 eV. In the spin
optimization, the initial spin configuration is taken as one net
spin on Co atoms. And the initial crystal structures of the
Ti$_{1-x}$Co$_{x}$O$_{2-\delta}$ with x=0.0625 and x=0.0417 are
taken as a (2$\times $2$\times $1) and a (2$\times $3$\times $1)
super-cell~\cite{19,20,21,22}, respectively, in which only one
apical Ti is replaced by Co. In the case of x=0.0625, one oxygen
atom can be removed from the octahedra containing Co or that
containing Ti, forming two different configurations called as the
third and fourth case, respectively. In every case, the
geometrical optimization is always made and convergence is assumed
when the forces on atoms are less than 50 meV/\AA, based upon
which the electronic structures are then calculated. Finally,
within the electric-dipole approximation, the imaginary parts of
the dielectric functions can be calculated by the following
formula:
\[
\begin{array}{l}
\varepsilon_2(\omega) =
\frac{8\pi^2e^2}{\omega^2m^2V}\sum\limits_{c,v}
{\sum\limits_k{\vert<c,k\vert{\rm{\bf\hat{e}}}\cdot }}{\rm{\bf
p}}\vert v,k
>\vert^2\times\delta[E_c(k)-E_v(k)-\hbar\omega],\\\end{array}\]\noindent
where $c$ and $v$ represent the conduction and valence band,
respectively. $\vert$$n, k>$ is the eigenstate obtained from
CASTEP~\cite{15} calculation. {\bf p} is the momentum operator and
${\rm {\bf \hat {e}}}$ is the external field vector. $\omega $ is
the frequency of incident photons. We use a 4$\times $4$\times $4
grid of k points to make an integration over the Brillouin Zone
(BZ). Using Kramers-Kr\"{o}nig (K-K) transformation, we can get
the real part $\varepsilon_1$ of the dielectric function and then
according to the formula of $\sigma = \sigma_1 + i\sigma_2 =
-i\frac{\omega}{4\pi}(\varepsilon_1 + i\varepsilon _2 - 1)$, the
optical conductivity is finally obtained.

\section{Results and Discussion} \label{Results and Discussion}
The CASTEP optimized geometrical structures together with the spin
densities in (010) plane of Ti$_{1 - x}$Co$_{x}$O$_{2-\delta}$ for
different x and $\delta$ values are shown in Fig. 1, in which the
number beside the atom represents the magnetic moment obtained by
Mulliken analysis, and its sign indicates the positive or negative
moment. It can be seen from Fig. 1 that the geometrical structure
of the system is changed very little by the doped Co atom, but one
oxygen atom vacancy causes a remarkable distortion of the original
octahedral structure, leading to changes of the electronic
structures. For example, comparison of Fig. 1a and 1b indicates
that more Co concentrations cause the spin density to be more
connective in space even without oxygen vacancy. On the other
hand, however, comparison of Fig. 1b and 1c shows that at the same
Co concentration, the geometrical structure distortion induced by
the oxygen vacancy will make the spin density further heavily
connected in space along the direction of Co --- Oxygen-vacancy,
indicating clearly that the oxygen vacancy plays a role of
enhancing the ferromagnetism. The emergency of spin density on the
Ti atom should be another noticeable indication of the heavier
connectivity of spin density in space than that without oxygen
vacancy at the same Co concentration. In Fig. 1(d) and 1(e),
although the oxygen vacancy is far away from the Co atom, it still
has larger effect on the physical properties of the system, which
is contrary to Park {\it et al.}'s result~\cite{19}. Some oxygen
atoms are pushed outside of the plane (010) in Fig. 1(d), and spin
density appears on Ti, making the spin density to be connective in
space too. Furthermore, we have also made an anti-ferromagnetic
(AFM) first principle calculation in the 3rd case (i.e, with
oxygen vacancy near Co atom), from which it is found that
independent of the initial spin configurations, the CASTEP
calculation performed in a doubly-enlarged supercell~\cite{20}
always converges to the FM ground state with an averaged magnetic
moment of about -0.9 $\mu_{B}$ on each Co atom, while for such a
supercell in the second case, the AFM ground state is always
obtained~\cite{20}, clarifying more clearly the importance of the
oxygen vacancy for FM coupling. More careful analyses show that
the electrons on Co and Ti are $d$-type like, and those on O atoms
are $p$-type like. And from Mulliken analysis, it is known that in
Fig. 1(c) the spin on Ti atom nearest to Co is anti-parallel to
that on Co, while the spin on the next-nearest Ti is parallel to
that of Co. In Fig.1(d), however, those spins on Ti and Co are all
in parallel. So the spin on Ti atom should not be ignored when we
try to pursue the origin of the ferromagnetism in the Co-doped
TiO$_{2}$. Comparison of Fig. 1(c) and 1(e) tells us that without
Co but only oxygen vacancy would not cause ferromagnetism.

\begin{figure}
\centering
\includegraphics[width=0.50\textwidth]{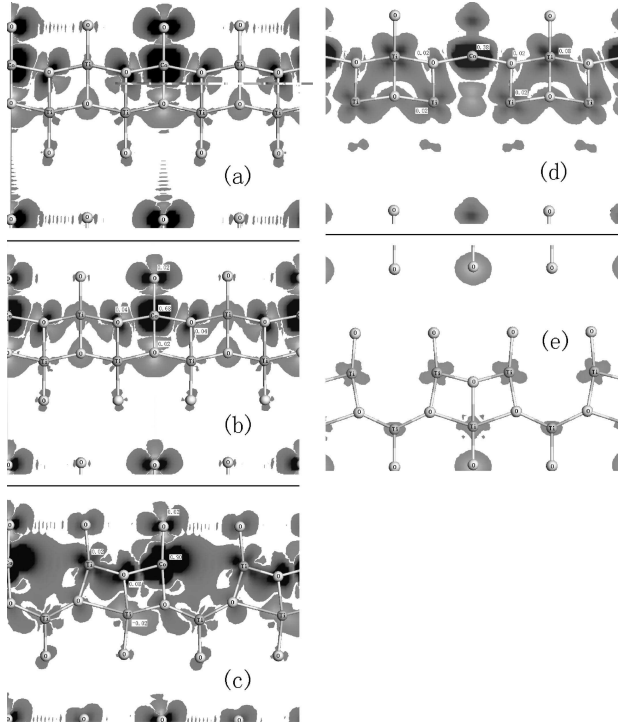}
\caption{The optimized structure and spin density of Ti$_{1 -
x}$Co$_{x}$O$_{2-\delta }$ in the plane (010) when (a) x=0.0417,
$\delta$=0; (b) x=0.0625, $\delta$=0; (c) x=0.0625,
$\delta$=0.0625 with oxygen vacancy near Co; (d) x=0.0625,
$\delta$=0.0625 with oxygen vacancy near Ti, which does not lie in
the plane (010); and (e) x=0.0625, $\delta$=0.0625 with oxygen
vacancy near Ti in the plane. The number beside the atom
represents its magnetic moment obtained by Mulliken analysis, and
its sign indicates the positive or negative moment.}\label{fig1}
\end{figure}

Now, we turn to the density of states (DOS) shown in Fig.2. Due to
the well-known shortcomings of the LSDA, our electronic structure
shows half-metallic Ti$_{1 - x}$Co$_{x}$O$_{2-\delta}$. It is
clearly seen from Fig. 2(a) and 2(b) that when x changes from
0.0417 to 0.0625, the DOS changes very little, not only the total
DOS but also all the partial DOSs for each kind of atoms. Compared
with the LMTO results obtained by Park {\it et al.}~\cite{19}, it
is known that the main difference lies in the energy region above
2 eV. On the other hand, the oxygen vacancy changes the DOS more,
especially those around the Fermi level. Fig. 2(c) shows that the
Fermi level shifts upward by about 1.2 eV compared to Fig. 2(b)and
the major spin becomes spin up. The oxygen vacancy near Co also
makes the first gap above Fermi energy narrower, from about 1 eV
to 0.5 eV, and thus will make the thermal excitation easier and
generate more charge carriers. As shown in Fig. 2(d), the oxygen
vacancy near Ti has bigger effects on the total DOS than near Co
shown in Fig. 2(c), shifting the Fermi level further up by about
0.75 eV, and causing much less DOS on Fermi level. Most
importantly, in this case, the system is no longer a half-metal
but a metal.

\begin{figure}
\centering
\includegraphics[width=0.5\textwidth]{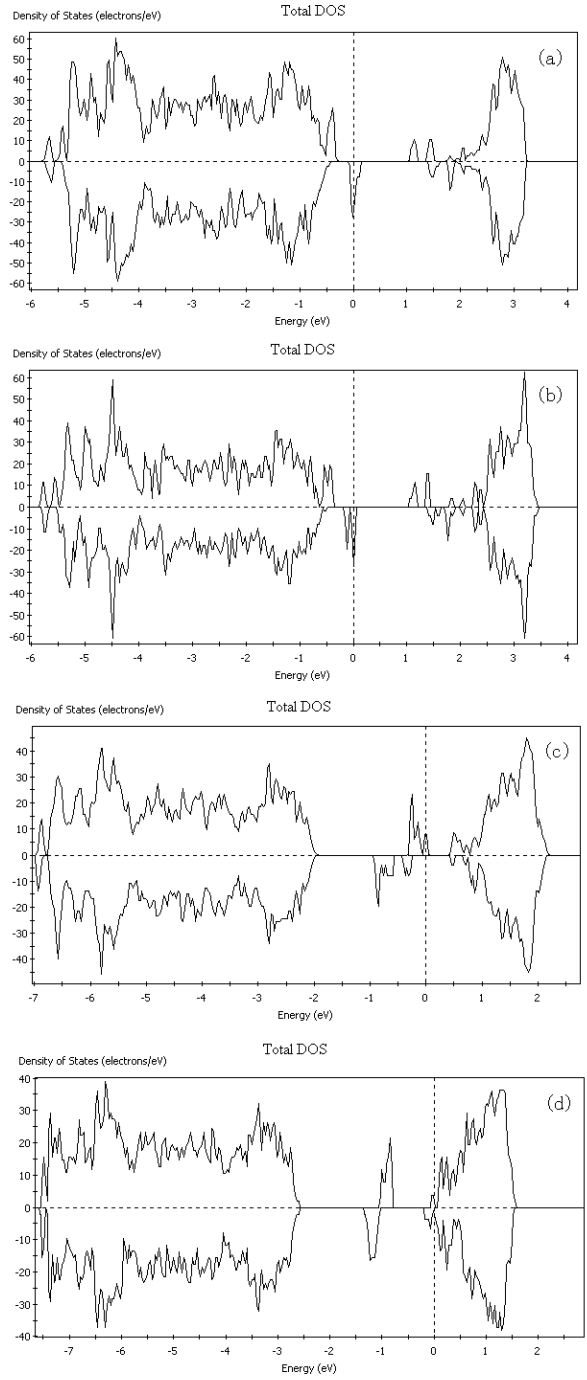}
\caption{The spin polarized DOS of Ti$_{1 - x}$Co$_{x}$O$_{2 -
\delta }$ when (a) x=0.0417, $\delta$=0; (b) x=0.0625, $\delta$=0;
(c) x=0.0625, $\delta$=0.0625 with oxygen vacancy near Co; and (d)
x=0.0625, $\delta$=0.0625 with oxygen vacancy near Ti. The
positive DOS means that of spin up while the negative is for spin
down.}\label{fig2}
\end{figure}

It is a powerful and natural method to investigate the electronic
structure of material by its optical response. The experimental
absorption edge of the Co-doped anatase TiO$_{2}$ is the same as
that of the pure anatase TiO$_{2}$, i.e., at about 3.2
eV~\cite{14}. So, in our calculation, we have used the scissor
approximation to fit the calculated absorption edge to the
experimental value, and shown in Fig. 3 thus obtained optical
conductivities for all four cases. For clarity, the spectrum in
the range of 0 $\sim$ 5 eV is magnified and shown in the inset. It
is seen from Fig.3(a) and 3(b) that in the energy range of 0
$\sim$ 6 eV, the peak values of x=0.0417 are smaller than those of
x=0.0625, which seems to be contrary to the experimental data in
Ref. [14] because there the corresponding peak values decrease
with x increasing. But, the peak positions in both cases match
well, indicating that different Co concentrations do not change
the peak positions. Again, we focus our attention to the oxygen
vacancy effects on the optical property of Ti$_{1 -
x}$Co$_{x}$O$_{2-\delta}$. It is seen from Fig. 3(c) that the
absorption edge shifts to a  higher energy at about 3.6 eV and its
first peak appears at about 4.0 eV, which is not shown in the
experiment because its measurement range is limited to 3.7 eV and
further experimental data are needed to confirm this point. While
above 6.0 eV, the peak positions, even their values, are the same
as those of the second case with no oxygen vacancy, which can be
explained by the difference between two DOSs in Fig. 2(b) and
2(c). As shown in Fig. 2(d), the oxygen vacancy near Ti has led to
completely new DOSs around the Fermi level, giving thus a rather
different optical responses, compared with other three cases. Now,
the first peak lies at about 3.5 eV, which is not seen in the
experiment. So, taking the total energy into account, we consider
that in the real materials, the oxygen vacancy rarely occupies the
site near Ti, but prefers near Co.

\begin{figure}
\centering
\includegraphics[width=0.5\textwidth]{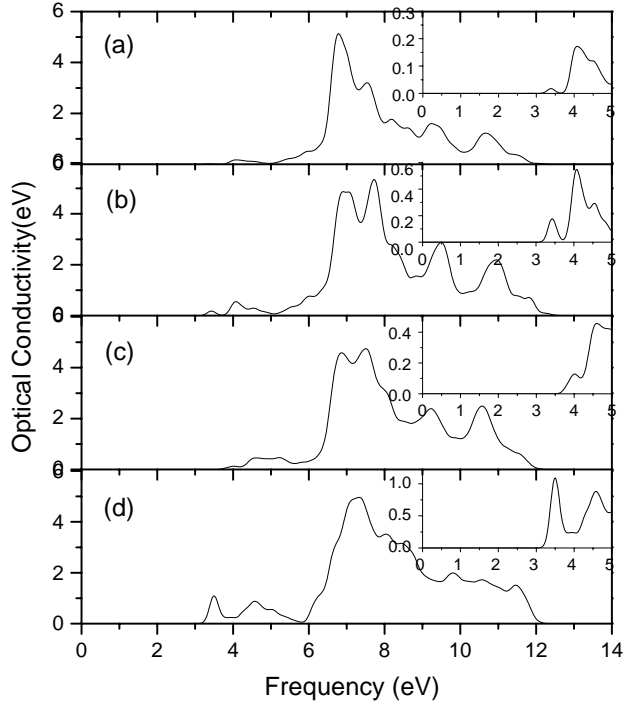}
\caption{The optical conductivity of Ti$_{1 - x}$Co$_{x}$O$_{2 -
\delta }$ when (a) x=0.0417, $\delta$=0; (b) x=0.0625, $\delta$=0;
(c) x=0.0625, $\delta$=0.0625 with oxygen vacancy near Co; and (d)
x=0.0625, $\delta$=0.0625 with oxygen vacancy near Ti. Each inset
is the magnified spectrum in the energy range between 0 and 5
eV.}\label{fig3}
\end{figure}

In order to overcome the shortcomings of the simple LSDA
calculations and also compare our calculated results more
reasonably with the experiments, we have further made the LSDA+U
calculations in the case of x=0.0625, $\delta$=0.0 in the
FP-LAPW\cite{23} formalism with U=3.0eV~\cite{19} by using the
same CASTEP optimized structure~\cite{20}. Obtained DOS in this
case is shown in Fig. 4(a). Comparing it with Fig. 2(b) obtained
by CASTEP, we can see that there is no big change of the DOS for
the most of the valence and conduction bands, except for
appearance of an energy gap $\approx$ 0.8 eV around the Fermi
level, indicating that the system becomes a semiconductor and is
in good agreement with the experimental data for the case of
x=0.0625, $\delta$=0.0. In addition, we have also calculated its
optical conductivity by the LSDA+U in the FP-LAPW formalism with
the same U=3.0 eV. Due to well known reason, the energy gap of
about 0.8 eV obtained by the FP-LAPW+U method is still much less
than the experimental value of $\sim$3.2 eV. Therefore, the same
scissor approximation is used to fit the absorption edge. The
obtained result is shown in Fig. 4(b), which, compared with Fig.
3(b), seems to match better the observed data, not only the peak
positions but also their heights, indicating that our
Co-substitutional model for the Ti$_{1 - x}$Co$_{x}$O$_{2-\delta}$
is rather reasonable and the on-site Coulomb interaction of the Co
atom plays an important role in the physical properties of the
system, and in general should be included.

\begin{figure}
\centering
\includegraphics[width=0.5\textwidth]{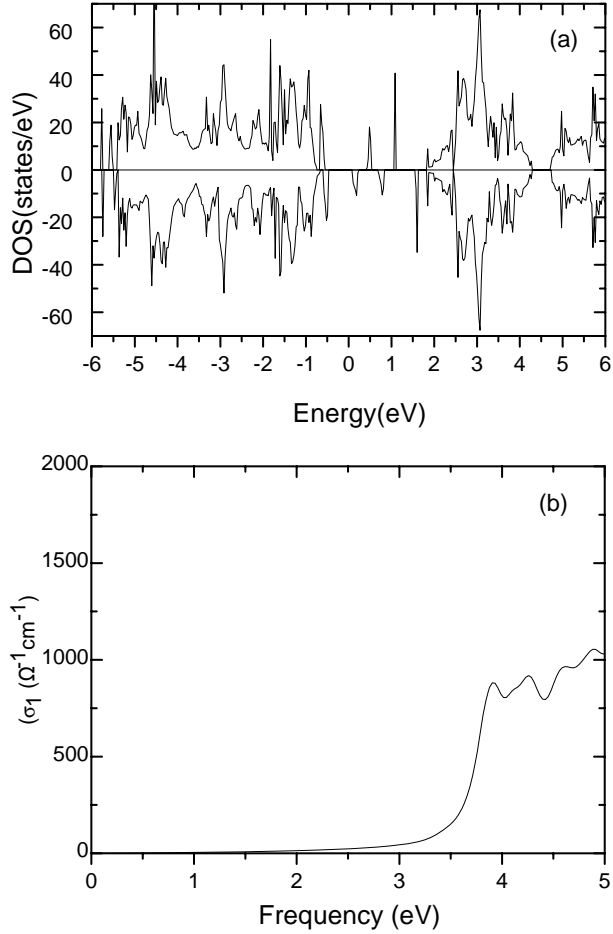}
\caption{The DOS(a) and optical conductivity(b) of Ti$_{1 -
x}$Co$_{x}$O$_{2 - \delta }$ for x=0.0625, $\delta$=0, calculated
by LSDA+U in the FP-LAPW formalism with U=3.0 eV.}\label{fig4}
\end{figure}

\section{Conclusion} \label{Conclusion}
In summary, assuming the substitutional doping of Co for Ti, we
have studied the effects of different Co-doping concentrations and
oxygen vacancies on the electronic structures and optical
properties of the Ti$_{1 - x}$Co$_{x}$O$_{2-\delta}$ by the first
principles calculations. Our obtained results show that the doped
Co-concentrations from 0.0417 to 0.0625 have only a little effect
on them, but the oxygen vacancy and its distribution in the system
have much larger influences, especially on the ferromagnetism and
optical properties of the Co doped anatase TiO$_2$.

\begin{acknowledgments}
The authors thank support to this work from a Grant for State Key
Program of China through Grant No. 1998061407 and National Science
Foundation under Grant No. 90103038. J. D. acknowledges greatly
support from a Grant under the Grant No. 14GS0204 in the IMR of
Tohoku university in Sendai of Japan and valuable discussions
about the Ti$_{1-x}$Co$_{x}$O$_{2-\delta}$ with Dr. T. Fukumura in
the IMR. Our calculations have been done on the SGI origin 2000
and 3800 Computers.
\end{acknowledgments}

\end{document}